\documentclass[%
 aip,
 jcp,%
 amsmath,amssymb,
 reprint,%
]{revtex4-2}

\usepackage{graphicx}
\usepackage{dcolumn}
\usepackage{bm}
\usepackage[mathlines]{lineno}
\usepackage{subfigure}
\usepackage{mhchem}
\usepackage{physics}
\usepackage{hyperref}
\hypersetup{
    colorlinks=true,
    linkcolor=blue,
    filecolor=magenta,      
    urlcolor=cyan,
    pdftitle={Draft Manuscript},
    pdfpagemode=FullScreen,
    }

\usepackage{xcolor}
\usepackage[normalem]{ulem}
\definecolor{erb}{RGB}{75, 0, 130}
\definecolor{bhavayedits}{RGB}{128, 0, 255}

\begin{document}


\title{Noise-induced synchronization in coupled quantum oscillators}

\author{Eric R. Bittner}
\affiliation{%
Department of Physics, University 
of Houston, Houston, Texas, 77204, United~States.
}%
\email{bittner@uh.edu}
\author{Bhavay Tyagi}
\affiliation{%
Department of Physics, University 
of Houston, Houston, Texas, 77204, United~States.
}%

\date{\today}

\begin{abstract}
We consider the 
quantum dynamics of 
a pair of coupled quantum oscillators coupled to a common correlated dissipative environment. The resulting 
equations of motion for both the operator
moments and covariances can be integrated
analytically using the Lyapunov equations.
We find that for fully correlated
and fully anti-correlated environments, 
the oscillators relax into a phase-synchronized state that persists for 
long-times when the two oscillators are 
nearly resonant and (essentially) forever
if the two oscillators are in resonance. 
We identify an exceptional point that indicates the
onset of broken symmetry between an unsynchronized 
and synchronized dynamical phase of the system 
as correlations within the environment are increased.
We also show that
the environmental noise correlation leads to quantum entanglement, and all the correlations between the two oscillators
are purely quantum mechanical in origin. This work provides a robust mathematical foundation for understanding how long-lived exciton coherences can be linked to vibronic correlation effects.
\end{abstract}

\maketitle

\section{Introduction}

While recovering from an illness in 1665, Christiaan Huygens observed that two identical pendulum clocks, when suspended from a common substantial beam, quickly synchronized with the same periods and amplitudes.~\cite{Huygens:1669aa,Huygens:1673aa,Ellicott:1739aa,Ellicott:1739ab,Airy:1830aa,Newton:1833aa,Korteweg:1906aa,huygens1916oeuvres}.
He described this intriguing behavior as ``an odd sympathy" and concluded that synchronization occurs due to the small vibrations transmitted through the support, which acts as a coupling mechanism between the pendula. Over time, these interactions facilitate an exchange of energy that synchronizes their motion. 
He presented these results to the Royal Society 
using a series of complex geometric arguments
to explain the synchronization mechanism\cite{huygens1916oeuvres,Huygens:1673aa}. 
Surprisingly, the problem of precise
time-keeping at sea remains a problem even today. 
While current laboratory-based atomic clocks reach an accuracy of 1 part in $10^{18}$ 
per day, having this standard at sea is made 
difficult by the spatial and environmental 
conditions imposed by a marine environment. 
A recent report by Roslund \textit{et al.}
indicate an accuracy of about 300ps per day over 20 days on a 
ship-borne device using the optical 
transitions of \ce{I2} vapor as their time standard. Such precision is needed to 
achieve nano-second synchronization of clocks
for making ultra-precise measurements, sub-meter scale GPS position finding, and long-distance space missions. 

The concept of 
synchronization has wide-ranging implications across various scientific fields, including the coordination of biological rhythms, the operation of power grids, and the development of synchronized communication systems~\cite{Strogatz:1993aa,Pikovsky:2001aa,Bennett:2002aa,Goychuk:2006aa,Oliveira:2015aa,Pena-Ramirez:2016aa}.
On the microscopic level, 
synchronization is crucial for
developing quantum technologies
that are robust against
environmental effects such as
decoherence
\cite{Chegnizadeh:2024aa}, as well as in
understanding the underlying
mechanisms for the transport
of quantum excitations in 
photosynthetic systems
in which observed long-lived exciton coherences are linked to vibronic correlation effects. The enhanced excitonic coherence has a profound influence on the enhancement of exciton transport and delocalization~\cite{engel2007evidence,panitchayangkoon2010long,chenu2015coherence,scholes2017using,Zhu:2024aa,Du:2021aa,Olbrich:2011aa,Wang:2015aa}.

We recently proposed that
synchronization between pairs of qubits could be 
enhanced by coupling  to a noisy ancillary 
environment with correlated or anti-correlated
coupling.\cite{PMC2024,JPCL2024} In these works,
we show that noise correlation can help to 
preserve the fidelity and purity of a prepared
state and determines the
degree of phase synchronization such that anti-correlated noise leads to anti-phase synchronization 
and correlated noise leads to in-phase synchronization. 
In this work, we consider a pair of coupled quantum 
oscillators coupled to a common correlated dissipative environment. We show that correlations
within the environment act as a shared 
resource that leads to entanglement in the steady-state solutions.  
The model is  applicable 
to a wide range of physical systems from coherent 
exciton transport in light-harvesting 
complexes to quantum phase transitions in driven systems. Hence, we conclude
that spontaneous synchronization 
is a universal hallmark of the interaction 
between a quantum system and a correlated environment. 

\section{Lyapunov equations}
We begin by writing 
a suitable Hamiltonian for this system
\begin{align}
    H/\hbar = \omega_1 a_1^\dagger a_1
    +\omega_2a_2^\dagger a_2
    +g (a_1^\dagger a_2 + a_2^\dagger a_1)
\end{align}
in which $[a_i,a_j^\dagger] = \delta_{ij}$ are
bosonic operators with harmonic frequencies
$\{\omega_i\}$ coupled via $g$ which allows
the exchange of quanta from one local mode to the 
other. For the isolated system, 
the total 
number of excitations is conserved
since $[a_1^\dagger a_1+a_2^\dagger a_2,(a_1^\dagger a_2 + a_2^\dagger a_1)]= 0$.
Further, we can remove the coupling 
term by transforming to a normal-mode representation
without loss of generality. 
However, for purposes of our analysis
we shall continue to work in a local-mode 
representation. 
The equations
of motion take the general form
\begin{align}
    \frac{d\rho}{dt} =
    -i[H,\rho] + \sum_{i}D_i(\rho)
\end{align}
where $D_i(\rho)$ is the total Lindblad dissipator for oscillator $i$ which 
has the general form
\begin{align}
    D_i(\rho) = L_i\rho L_i^\dagger - \frac{1}{2}\{L_i^\dagger L_i, \rho\} + L'_i\rho {L'}_i^\dagger - \frac{1}{2}\{{L'}_{i}^{\dagger} L'_i, \rho\}
\end{align}
where $L_i$ and $L'_i$ are Lindblad operators describing the local 
system/bath interactions associated with oscillator $i$.\footnote{We note that our analysis 
was done 
within Mathematica (v14.x)
by defining a series 
bosonic algebra 
rules to generate and simplify the
rather cumbersome expressions
for multi-component boson systems. Included in this 
is a similar set of rules and 
transformations useful for 
multi-components spin systems.
We include a link to a notebook with our derivations in the Data Availability section. 
} 
For these, we use 
\begin{align}
    L_{i}&=\sqrt{\gamma (\overline{n}_i+1)}a_i \\
     L'_{i}&=\sqrt{\gamma \overline{n}_i}a^\dagger_i 
\end{align}
for the local Lindblad terms
where $\gamma$ give the thermalization
rate and $\overline{n}_i$ is the 
thermal population.

Following from our previous work, and assuming that the 
coupling to a common environment can be 
described by correlated Wiener processes, the correlation 
between the dissipative channels can be 
introduced by defining a correlation matrix, $\Xi$
which takes the form
\begin{align}
    \Xi = \left(
\begin{array}{cc}
    1 &\xi  \\
    \xi  & 1 
\end{array}
    \right)
\end{align}
 in which $-1\le \xi\le 1$ serves as a correlation 
 parameter. 
 In general, the
correlation matrix $\Xi$ describes the graphical adjacency
between Wiener processes such that 
$dW_i(t)dW_j(t') = \Xi_{ij}\delta(t-t')dt$.

 We use the eigenvalues 
 and corresponding eigenvectors
 of $\Xi$ to 
 construct new Lindblad 
 operators 
 \begin{align}
     L_\alpha = \sqrt{\xi_\alpha}\sum_{i}T_{\alpha i}L_i 
 \end{align}
 where  $T^\dagger \Xi T = {\rm diag}\{\xi_\alpha\}$.
In general, the
correlation matrix, $\Xi$, describes the 
topological adjacency between correlated
Wiener processes such that 
$dW_i(t)dW_j(t') = \Xi_{ij}\delta(t-t')dt$
and may represent a connected graph 
belonging to a particular discrete symmetry 
group.
Consequently, the eigenvectors of $\Xi$ can
be classified according to the irreducible
representations of that group.  
For the special case of a $k$-regular graph (that is a graph with 
each node having $k$ edges 
and $\Xi = I + \xi A$, where $A$ is the adjacency matrix for an undirected graph and $I $ is the identity matrix), one can always find one eigenstate of $\Xi$ with an eigenvalue 
equal to 0.
For example, 
for a cyclic 2-regular
graph with $N$ nodes
\begin{align}
    \xi_\alpha 
    = 1 + 2 \xi \cos(\frac{2\pi}{N}\alpha
    )
\end{align}
where $\alpha = 0,1, \lfloor N/2 \rfloor)$. 
The eigenvectors
with $1\le \alpha \le \lceil N/2 \rceil-1$ are doubly degenerate.
For even $N$, 
the $\alpha = 0$ 
and $\alpha = N/2$
eigenvalues are
$0$ and $2\xi$, 
respectively.
with the 
corresponding 
eigenvectors
are $(1,1\cdots,1)$
and $(1,-1,1,-1\cdots)$.
However, if $\xi=1$, the $\alpha = 0$ eigenvector is the totally anti-symmetric eigenvector
and if $\xi = -1$, then the $\alpha = 0$ eigenvector is totally symmetric.
Consequently, the 
symmetry of the 
$0$-mode
implies that corresponding irreducible representations of the system will be totally 
{\em decoupled} from the environment.

  
In our calculations which follow, we use the following set
of local Lindblad operators:
\begin{align}
    L_i = \sqrt{\gamma (\overline{n}_i+1)}a_i \\
    L_{i}' = \sqrt{\gamma \overline{n}_i}a^\dagger_i
\end{align}
where $L_i$ describes the relaxation and $L'_i$ describes the thermal excitations. The relaxation
rate is specified by $\gamma$ and $\overline{n}_i$ is
the equilibrium thermal population of oscillator $i$.
 We use the eigenvalues 
 and corresponding eigenvectors
 of $\Xi$ to 
 construct new Lindblad 
 operators corresponding to 
 the correlated and anticorrelated
 components 
 \begin{align}
     L_\pm = \sqrt{\frac{1\pm \xi}{2}}(L_1 \pm L_2)\\
     L'_{\pm} = \sqrt{\frac{1\pm \xi}{2}}(L'_1 \pm L'_2).
 \end{align}
 The total dissipator now includes cross terms involving $L_1$ and $L_2$ and similarly $L'_1$ and $L'_2$ arising from correlation between the two local environments.
 When $\xi = 0$, the cross components
 vanish and dissipation channels are independent of each other. When $\xi=1$, the 
 anti-symmetric contribution vanishes
 leaving only symmetric terms $L_+$ and $L'_+$. Likewise, in case of anti-correlated noise ($\xi = -1$), symmetric terms vanish leaving only the anti-symmetric terms $L_-$ and $L'_-$.

In principle, one can also write $H$ in terms of the normal modes of the oscillators themselves. For the the two oscillator system we have a symmetric and an anti-symmetric normal mode.
Consequently when $\xi>0$ (more correlated environment), the symmetric normal mode will be strongly damped compared to the anti-symmetric mode such that in the extreme case of $\xi <0$ the anti-symmetric mode is completely undamped and vice-versa for $\xi =0$.


Since our equations of motion generate a Gaussian map, we will instead work with the operator expectation values and covariances rather than 
with the density matrix itself.
To establish our notation, we  write the vector of moments of the system operators as  ${\bf x} = \{\langle a_1 \rangle,
\langle a^\dagger_1 \rangle,
\langle a_2\rangle, 
\langle a_2^\dagger \rangle\}$
and the
covariances as
\begin{align}
    \Theta_{ij} = \frac{1}{2}\langle \{\delta x_i^{\dagger},\delta x_j\}\rangle = 
    \frac{1}{2}
    \langle \{x_i^{\dagger},x_j\}\rangle - \langle x^{\dagger}_i\rangle
    \langle x_j \rangle
\end{align}

The covariance matrix, $\Theta$, must be 
positive, semi-definite and should satisfy
the uncertainty principle at all times.  This can be 
stated compactly 
as 
\begin{align}
    \Theta + i\Omega \ge 0
\end{align}
where $\Omega$ is the symplectic form
\begin{align}
    \Omega = \oplus_{n=1}^N \left(\begin{array}{cc}0 & 1 \\ -1 & 0 \end{array} \right).
\end{align}
For the case at hand, we can see that 
eigenvalues of $\Theta$ are always positive, semi-definite.  Likewise, the symplectic form 
remains invariant under any symplectic transformation, 
including the one that diagonalizes $\Theta$
(Williamson's Theorem)\cite{williamson1936algebraic}.
Consequently, $\Theta$ represents a
\textit{bona fide} covariance matrix for a physical 
Gaussian state~\cite{PhysRevLett.109.190502}.

The covariance matrix and moments can be used to 
construct the Wigner function in multi-variable form
\begin{align}
    {\cal{W}}(\alpha) = 
    \frac{1}{(2\pi)^N\sqrt{|\Theta|}}\exp\left[-\frac{1}{2}
    (\alpha-{\bf x})^\dagger\cdot\Theta^{-1}\cdot(\alpha-{\bf x}).
    \right]
\end{align}
where $N$ is the number of oscillators and $\alpha$ is a continuous multi-dimensional complex variable whose real and imaginary components are related to the position and momentum conjugate variables. 
Note, that the Husimi-Q function and Glauber-Sudarshan P functions
are similarly constructed; however, the Q-function has a covariance $\Theta_Q = \Theta + I$ while the P-function uses $\Theta_P = \Theta-I$.

The time-evolution of any observable, $\hat O$, is
given by the master equation
\begin{align}
    \frac{d\langle \hat O\rangle}{dt} = 
    i\langle [H,\hat O] \rangle + \langle \overline{D}_L
    (\hat O)\rangle
\end{align}

Since these are equations 
of motion for the expectation 
values rather than the operators
themselves, we use the adjoint 
dissipator
\begin{align}
    \overline{D}_L(\hat O) = \langle \hat{O} D_L(\rho)\rangle &= {\rm tr}\left[
    {\hat O} \left(L\rho L^\dagger - \frac{1}{2}\{L^\dagger L,\rho\}\right)
    \right] \nonumber \\
    &=
    \frac{1}{2}
    \langle
    L[\hat O,L^\dagger]+
    [L,\hat O]L^\dagger\rangle
\end{align}
in deriving these terms.
\begin{widetext}

It is straightforward to show that the moments, $\bf x$, and 
covariance, $\Theta$,
evolve according to 
the following set of equations of motion
\begin{align}
    \frac{d{\bf x}}{dt} = W\cdot {\bf x} + f
\end{align}
where $W$ is the  dynamical 
matrix and $f$ includes any 
source or sink terms.
Similarly the covariances 
evolve according to
\begin{align}
    \frac{d\Theta}{dt}
    = W \cdot \Theta +\Theta \cdot W^\dagger
    + F
    \label{eq:8}
\end{align}
where matrix $F$ depends only 
on the dissipative terms in the
Liouville equation. 
These are referred to as the 
Lyapunov equations and 
are useful for understanding the 
stability and dynamics 
of chaotic systems\cite{PARKS:1992aa}.

For the coupled oscillators, we find the dynamical matrix is 
given by 
\begin{align}
\label{eq:17}
    W
    = \left(
\begin{array}{cccc}
    -i\omega_1 & 0          & -ig -\xi \gamma_{12}/2& 0   \\
     0         & +i\omega_1 & 0    & ig-\xi\gamma_{12}/2 \\
     -ig  -\xi\gamma_{12}/2      & 0         & -i\omega_2  & 0 \\
     0         & ig-\xi\gamma_{12}/2         &   0        & +i\omega_2
\end{array}
    \right)  -\frac{\gamma}{2}I_4
\end{align}
where $I_4$ is the $4\times 4$ identity matrix and
 \begin{align}
     \gamma_{12} &=\gamma( \sqrt{(\bar{n}_1+1)(\bar{n}_2+1)}
     -\sqrt{\bar{n}_1 \bar{n}_2})
\end{align} is a temperature
dependent relaxation rate.
Similarly,
\begin{align}
    F = 
   - \left(
\begin{array}{cccc}
 \frac{\gamma}{2}+\gamma \overline{n}_1 & 0 & \xi  \gamma\sqrt{\overline{n}_1 \overline{n}_2} & 0 \\
 0 & \frac{\gamma}{2}+\gamma \overline{n}_1 & 0 & \xi  \gamma\sqrt{ \overline{n}_1 \overline{n}_2} \\
 \xi \gamma \sqrt{\overline{n}_1 \overline{n}_2} & 0 & \frac{\gamma}{2}+\gamma \overline{n}_2 & 0 \\
 0 & \xi \gamma \sqrt{\overline{n}_1 \overline{n}_2} & 0 & \frac{\gamma}{2}+\gamma \overline{n}_2 \\
\end{array}
\right)
\end{align}

The derivation of these equations 
is straightforward and was facilitated by a Mathematica package we developed for doing multi-boson operator algebra. This package is publically available from the URL given in the Data Availability Statement below.

\end{widetext}
Both $W$ and $F$ can be rearranged into block-diagonal form and from the Lyuapunov equations, we obtain the 
equations of motion for the moments and covariances.


The dynamical matrix, $W$, is non-Hermitian and 
has complex eigenvalues
\begin{align}
\label{20}
    \lambda_\pm
    =\left(\frac{i}{2}(\omega_1+\omega_2)-\gamma
    \right)
    \pm
\frac{i}{2}    \sqrt{
(2g -i\xi \gamma_{12})^2
+(\omega_1-\omega_2)^2
    }
\end{align}
If we take the case where
the two oscillators are decoupled ($g=0$), 
synchronization becomes 
spontaneous when 
$\xi > (\omega_1-\omega_2)/\gamma_{12}$.

This defines the critical value of
$\xi$ necessary for synchronization 
to occur.  As $\xi\to\pm 1$, one of the two normal modes will become 
completely decoupled from the 
dissipative environment, leading to
the formation of undamped oscllations
in which the two oscillators are completely in-phase or out-of-phase with each other. 
This is the synchronized (or antisynchronized)
regimes.

\begin{figure*}
    \centering
    \subfigure[]{
   \includegraphics[width=0.45\linewidth]{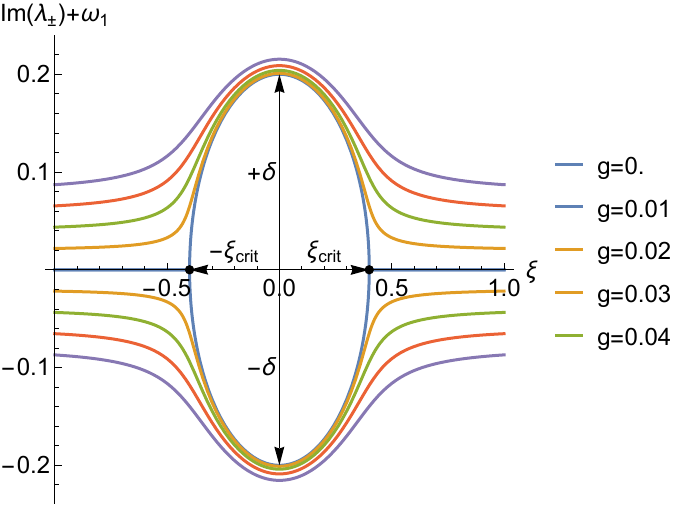}}
\subfigure[]{\includegraphics[width=0.45\linewidth]{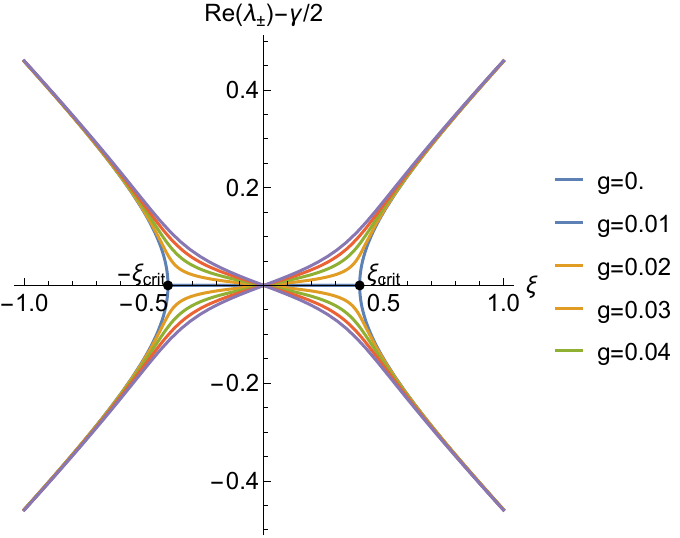}}
\subfigure[]{\includegraphics[width=0.45\linewidth]{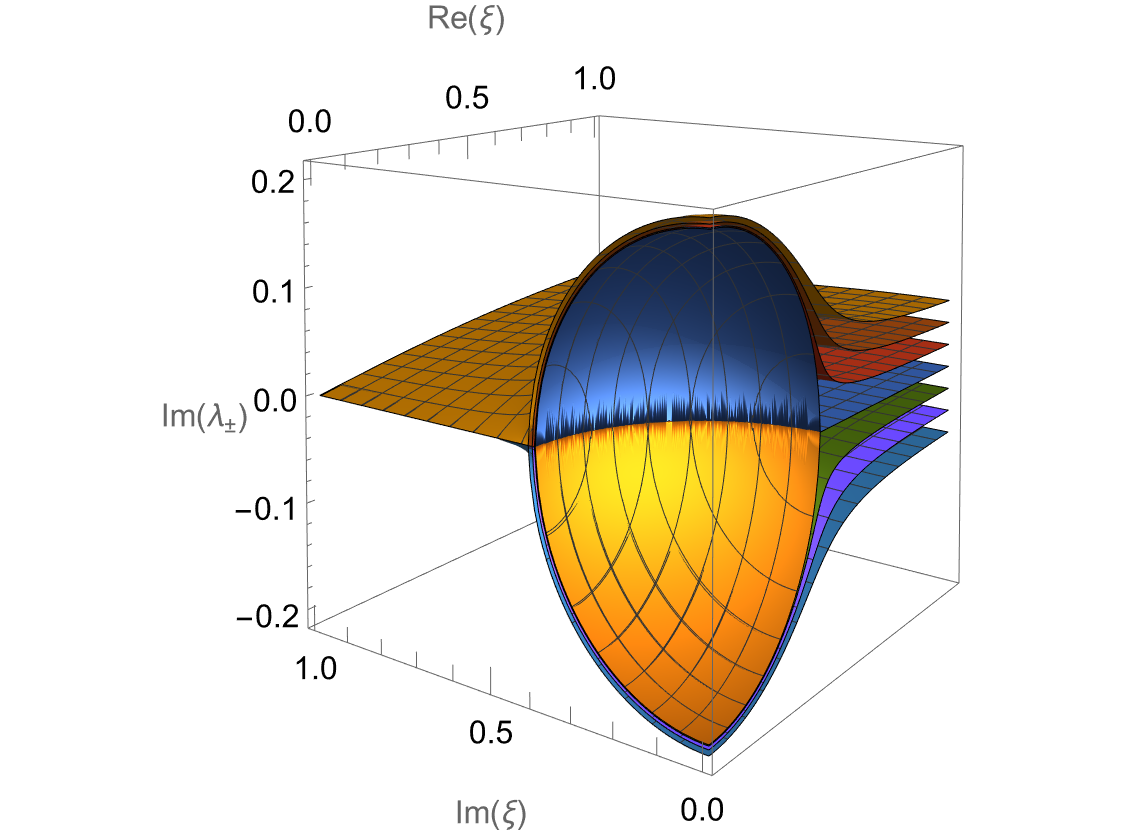}}
\subfigure[]{\includegraphics[width=0.45\linewidth]{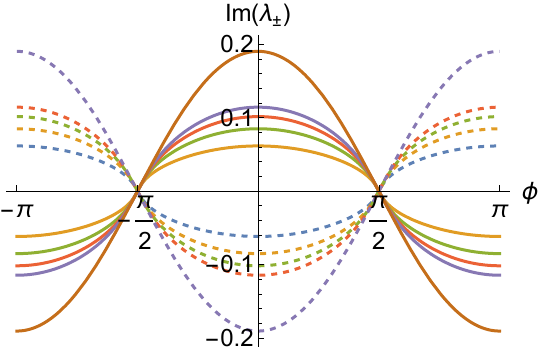}}

\caption{Real (a) and Imaginary (b) components of the eigenvalues of the non-Hermitian 
dynamical matrix in Eq.~\ref{eq:17}
for various 
parametric values:
$\delta = \omega_1-\omega_2 = 0.2$, $\gamma = 0.5$ and 
increasing values of 
exchange coupling
$g = 0$ to $g = 0.04$
The system enters the spontaneous synchronized domain above the critical coupling $\xi_{crit} = \delta/\gamma$.
(c) Analytical continuation of (a)
onto the complex $\xi$-plane 
for
showing the presence of a branch 
cut starting at $\pm i\xi_{crit}$.
(d) Slice through (c) along $|\xi_{crit}|$ passing through the 
exceptional points at $\phi =\pm \pi/2$ for
$g>0$.  Solid: $\text{Im}(\lambda_+)$. Dashed: $\text{Im}(\lambda_-)$.
}
    \label{fig:1}
\end{figure*}

Figure ~\ref{fig:1}(a,b) shows the real and
imaginary eigenvalues of the dynamical matrix for a system at $T=0K$
with detuning $\delta = 0.2$ and relaxation rate $\gamma = 0.5$. Here,  we set $\omega_1 = 1$ and $\omega_2 = \omega_1+\delta$ to set the time scale of our system. 
Both the real and imaginary components
of the eigenvalues are shifted as indicated on the plots. 
Here, the eigenvalues of $W$ appear to 
exhibit an exceptional point at $\xi_{crit}=(\delta/\gamma_{12})$, where
the two frequencies coalesce 
($\text{Im}(\lambda_+) = \text{Im}(\lambda_{-}$))
and one mode becomes less damped while the other becomes more strongly damped.
An exceptional point
signals  a breaking of 
 parity-time (${\cal P}{\cal T}$) symmetry.
\cite{bender1998real, bender2003must, bender2002complex, bender2007making, bender2002generalized, mostafazadeh2003exact, bagchi2000sl}
  In the normal regime, the system exhibits a balanced distribution of gain and loss. When this balance is disrupted, the system can transition into a phase where the ${\cal P}{\cal T}$ symmetry is broken. 
In a ${\cal P}{\cal T}$-symmetry broken system, the fluctuations can become asymmetric, meaning the probability of a fluctuation in one direction can be significantly different from the probability of a fluctuation in the opposite direction, leading to a violation of the fluctuation dissipation theorem.\cite{Marinari_1998}

Setting $g\ne 0$
lifts the degeneracy
between the two 
oscillators and there is no longer an exceptional point along the 
real $\xi$-axis.
However, one can find the exceptional point by performing
an analytical continuation 
of $\xi\to |\xi|e^{i\phi}$ in $W$.
This 
places the  exceptional point
along the
imaginary $\xi$-axis ($\phi = \pi/2$)
for all values of
$g$.  Further, this is the start of 
a seam  where $\text{Im}(\lambda_+) = 
\text{Im}(\lambda_-)$ along the imaginary
axis staring at $\pm i\xi_{crit}$.
This is illustrated
in Fig.~\ref{fig:1}(c) where we show 
the analytical continuation of
 $\text{Im}(\lambda_\pm)$ on the complex $\xi$-plane.
 Consequently, one can 
adiabatically transition between 
synchronized and anti-synchronized 
regimes by transforming
the system along $|\xi| e^{i \phi}$.
While this may not be feasible in a material system, it may be possible 
to construct light-matter or fully
optical systems that exhibit this
behavior.\cite{Mukamel:2023aa} 
Lastly, we note that
$g<0$ swaps the two regimes 
such that $\xi<0$ leads to anti-synchronized and $\xi>0$ produces
synchronized dynamics.

\begin{figure*}
    \centering
   \subfigure[]{ \includegraphics[width=0.3\linewidth]{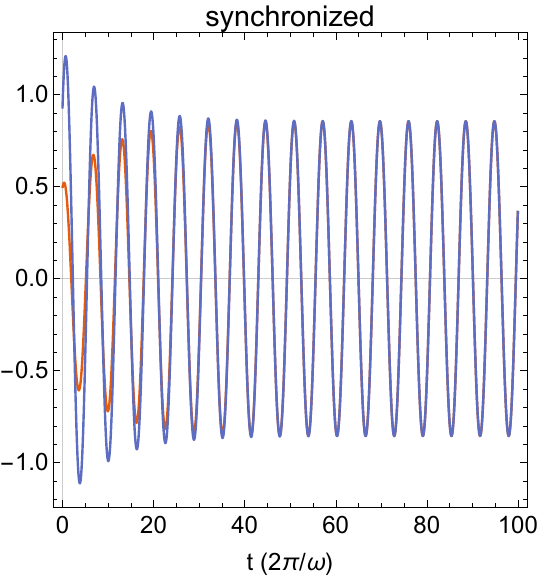}
}
   \subfigure[]{ \includegraphics[width=0.3\linewidth]{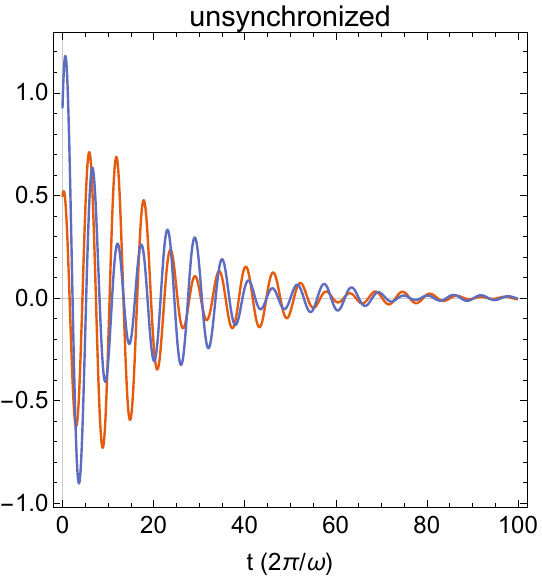}
}
\subfigure[]{
\includegraphics[width=0.3\linewidth]{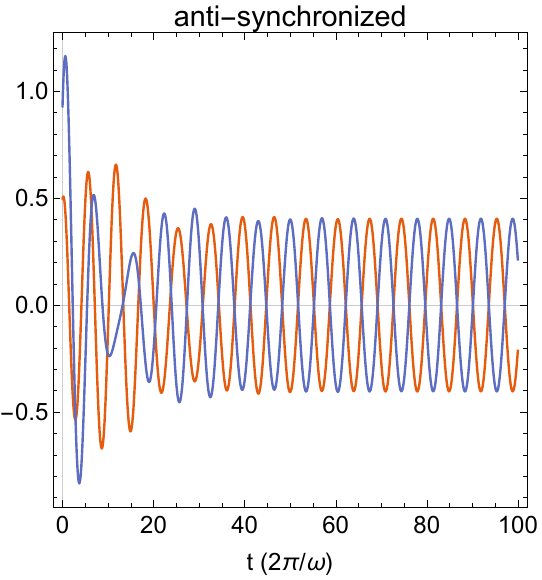}}
    \caption{Representative 
    moment trajectories
    for 
      pairs of equivalent}
    oscillators driven by 
    (a)anti-correlated $(\xi = -1)$, 
    (b) uncorrelated $(\xi = 0)$, 
    and (c) correlated ($\xi = +1)$ noise.
    (Parameters: $\delta = 0$, $g = 0.1$, $\gamma = 0.1$.)
    
    \label{fig:2}
\end{figure*}

In Fig.~\ref{fig:2} we show two representative trajectories for $\langle a_1(t)\rangle$ and $\langle a_2(t)\rangle$ 
for the case of zero detuning ($\delta = 0$)
and with various degrees of environmental correlations.
Here, we simply 
chose random initial conditions
for the $\langle a_1\rangle$ 
and $\langle a_2\rangle$ moments.
We see clearly that 
anti-correlated noise ($\xi = -1$) leads to in-phase synchronization while fully correlated
noise ($\xi = +1$) leads to anti-phase synchronization. Further more, at long time, the  synchronized states are fully decoupled from the environment and do not decay as compared to the fully uncorrelated case which 
relaxes to zero on the time-scale dictated by 
$\gamma_{12}$.  
Similar regimes have
been suggested by a series of  recent papers 
concerning harmonic chains coupled by dissipative nearest neighbor terms \cite{Wachtler:2020,Wachtler:2023,Wachtler:2024,moreno2023}.

\begin{figure*}
    \centering
  \subfigure[]{  \includegraphics[width=0.45\linewidth]{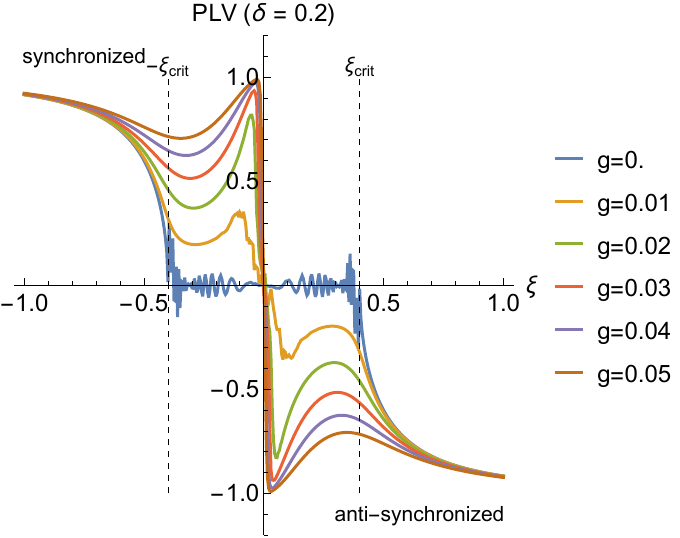}}
    \subfigure[]{  \includegraphics[width=0.45\linewidth]{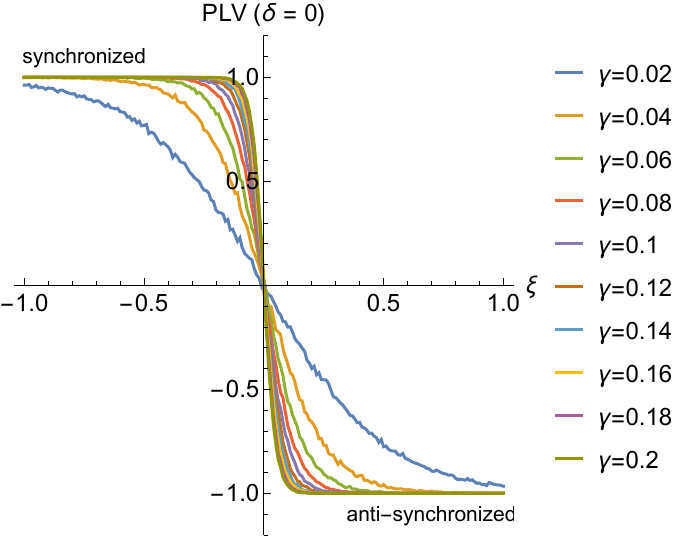}}
    \caption{(a) Average phase-locking value (PLV) 
versus $\xi$ for $\delta = 0.2$ and off-diagonal coupling $0\le g \le 0.05$.
(b)Average PLV for $\delta = 0$ and
$g=0$, varying the relaxation constant
$\gamma$.
}
    \label{fig:2a-new}
\end{figure*}

We
can quantify phase synchronization
by examining the 
phase-locking value (PLV), taking as the average phase difference
between $\langle a_1(t)\rangle$
and $\langle a_2(t)\rangle$,
\begin{align}
    PLV =
    \lim_{t_2\to\infty}
    \frac{1}{t_2-t_1}
    {\rm Re}\left\langle\int_{t_1}^{t_2}
    e^{i(\phi_1(t)-\phi_2(t))}dt
    \right\rangle
\end{align}
where $\phi_i(t) = \arg(\langle a_i(t)\rangle)$ is the 
phase-angle associated with the complex-valued moment.
For this, 
we first integrate the equation of motion for some sufficiently long
time, $t_1 \gg 1/\gamma$. We then 
sampled over 5000 initial phase differences, varying $g$  to accumulate the
data shown in Fig.~\ref{fig:2a-new}(a). 
For $g = 0$, synchronization does not
occur for  $-\xi_{crit} \le \xi \le \xi_{crit}$.  Outside this regime, 
the PLV goes to $+1$ for $-\xi_{crit} \geq -1$ and to $-1$ for $\xi_{crit} \leq +1$, indicating the presence of synchronized or anti-synchronized states. 
For $g\ne 0$, phase synchronization 
or anti-synchronization readily 
occurs, except when $\xi = 0$.
For $-\xi_{crit}\le \xi \le +\xi_{crit}$ and $g = 0$, 
the two oscillators are asymptotically
unsynchronized.  The oscilations in the computed PLV values are due to  numerical sampling of the initial 
conditions and 
finite time integration of
the PLV integral. These can be attributed to the competition 
between gain and loss mechanisms
while the system is in the 
non-synchronized regime. 

When the coupling $g \ne 0$, the PLVs show obvious dips 
close to the critical value of 
the correlation parameter indicating the ``competition'' between relaxation
and synchronization. 
In Fig~\ref{fig:2a-new}(b) we set $\delta = 0$ and vary the relaxation rate 
$\gamma$ to illustrate how the 
switch between synchronized to 
anti-synchronized behavior depends 
on the coupling to the environment, 
again averaging over initial conditions of the moments. 
Weakening the system/bath coupling
$\gamma$ weakens the synchronization 
effects.

\begin{widetext}
\subsection{Steady State Solutions}
It is 
useful to examine the long-time or stationary
state solutions (denoted by ``tilde") in which 
\begin{align}
    W\cdot \tilde\Theta + \tilde\Theta\cdot W^\dagger = -F.
\end{align}
The Lyapunov equation has a closed form 
solution for $\tilde\Theta$ when $\omega_{1} = \omega_{2}$
\cite{reference.wolfram_2024_lyapunovsolve},
\begin{align}
    \tilde\Theta_{11} &=\langle a_1^\dagger a_1\rangle_{ss} + 1/2 = 
    (\overline{n}_1 + \frac{1}{2})
    + \frac{2 g^2}{4 g^2 + \gamma^2}(\overline{n}_2-\overline{n}_1)
    + \xi^2\gamma_{12}\Delta
    \\
       \tilde \Theta_{22} &=\langle a_2^\dagger a_2\rangle_{ss} + 1/2 = 
    (\overline{n}_2 + \frac{1}{2})
    - \frac{2 g^2}{4 g^2 + \gamma^2}(\overline{n}_2-\overline{n}_1) 
    +\xi^2\gamma_{12}\Delta
    \\
    \tilde\Theta_{12}&= \langle a_1^\dagger a_2\rangle_{ss}= 
    i \frac{g\gamma}{4 g^2 + \gamma^2}(\overline{n}_2-\overline{n}_1)
    +\gamma \xi \Delta 
\end{align}
\end{widetext}
where $\Delta$ is a collection of variables
given by
\begin{align}
\Delta =   \frac{ \left(\gamma _{12} \left(\overline{n}_1+
\overline{n}_2+1\right)+2 \gamma\sqrt{
\overline{n}_1 \overline{n}_2}\right)}{2 \left(\gamma ^2-\gamma _{12}^2 \xi ^2\right)}.
\end{align}

This term imposes an additional 
constraint on the range of $\xi$ since it becomes singular when $\overline{n}_1\ne \overline{n_2}$
\begin{align}
 \xi = \pm\sqrt{\frac{1}{\overline{n}_1+\overline{n}_2+1}},
 \label{eq:19}
\end{align}
otherwise, $-1\le\xi\le 1$.  In what follows, we focus our attention on the special case where the two
oscillators are at the same frequency but may be held at different temperatures. 

\subsection{Correlation under thermal bias}
The off-diagonal $\tilde \Theta_{12}$ terms corresponds to the steady-state
 ``flow'' of quanta between the two oscillators. 
Interestingly, if there is a
 temperature bias between 
 the two oscillators, the system develops correlation since
 $\tilde\Theta_{12}\ne 0$. However, a temperature bias does not create synchronisation (or anti-synchronisation) since the sign of $\gamma_{12}$ in the equations of motion for the moments does not change if $\bar{n}_1 \neq \bar{n}_2$.
 In fact, if we  consider the 
 time evolution of the population of oscillators 1 and 2 at different temperatures:
 \begin{align}
     \frac{d n_1(t)}{dt} = \gamma(\overline{n}_1-n_1(t))
     - i g (\langle a_1^\dagger a_2\rangle + \langle a_2^\dagger a_1\rangle)\\
     \frac{d n_2(t)}{dt} = \gamma(\overline{n}_2-n_2(t))
     + i g (\langle a_1^\dagger a_2\rangle + \langle a_2^\dagger a_1\rangle).
 \end{align}
 For the stationary state (in which populations do not evolve with time), the two time-derivatives must be equal
and opposite in sign, and we can define a flux
 \begin{align}
      J = \frac{2 g \gamma }{4 g^2 + \gamma^2}(\overline{n}_2-\overline{n}_1)
 \end{align}
describing the flow of quanta from the warmer to cooler 
baths. 
This does not depend upon 
 correlations within the 
environment and only depends
upon the exchange coupling term within the Hamiltonian and the local system/bath coupling described by $\gamma$.

\subsection{Information Theory for Gaussian States}

\begin{figure*}
    \subfigure[]{\includegraphics[width=\columnwidth]{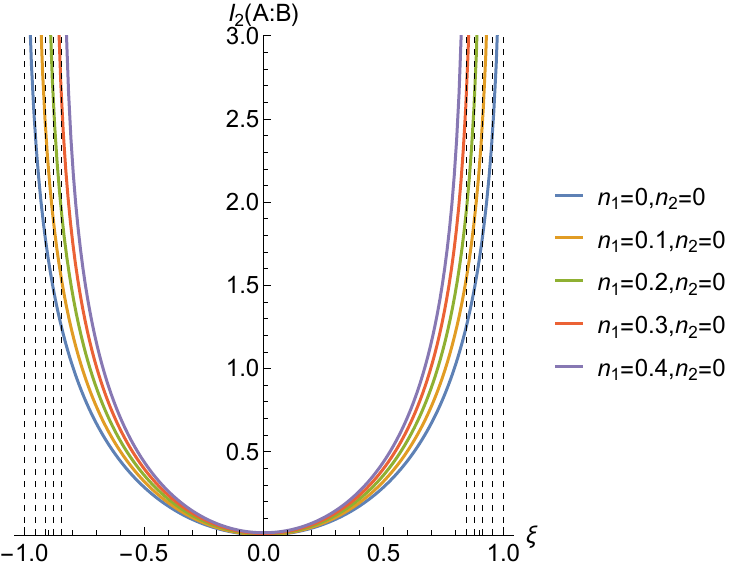}}
    \subfigure[]{\includegraphics[width=\columnwidth]{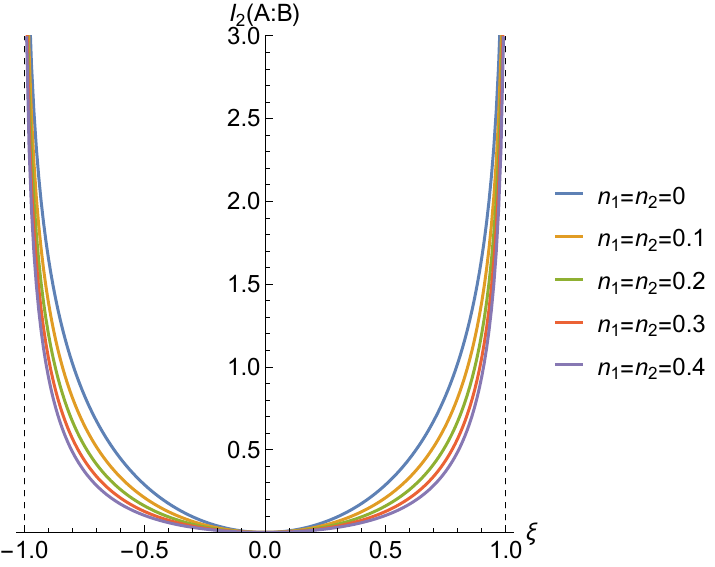}}
    \caption{(a) Mutual information 
    shared by two oscillators in the
    presence of a temperature bias as specified by the average occupations $\overline{n}_1$ and $\overline{n}_2$.
    (b) Mutual information shared by two oscillators at a common temperature. 
    In each case, we set $\omega_1=\omega_2 = 1$, $g = 1$, and $\gamma = 0.1$. 
    The vertical dashed lines correspond to 
    the limiting conditions imposed by Eq.\ref{eq:19} when $\overline{n}_1\ne \overline{n}_2$.}
    \label{fig:3}
\end{figure*}

The total entropy of a
bi-partite system 
is strongly sub-additive in that
$S_A+S_B \ge S_{AB}$.
This holds true for 
the von Neumann entropy
but is not true in general for the 
various orders of the
R{\'e}nyi
entropies. However,
strong sub-additivity for R\'enyi order 2 does 
hold for Gaussian states and is simply related to the 
purity\cite{PhysRevLett.109.190502}
\begin{align}
    S_2 &=  -\ln{{\rm tr}(\rho^2)}\\
    &=
    -\ln\int d\alpha
    {\cal W}(\alpha)^2.
\end{align}
One can easily show that for the 2-mode case,
\begin{align}
    S_2(\Theta) &= \frac{1}{2}\ln{{\rm det}\Theta}
    + N\ln{2}.
\end{align}
Furthermore, since the 
joint covariance matrix
can be written in 
block form as
\begin{align}
    \Theta = \left[ 
\begin{array}{cc}
    \Theta_A & \Gamma \\
     \Gamma^\dagger&\Theta_{B} 
\end{array}
    \right]
    \label{eq:25}
\end{align}
such that the covariance 
within subspace A or B
is given by the two 
diagonal blocks and the 
correlations between the two  are the 
off-diagonal blocks ($\Gamma$).
The mutual information
is then simply
\begin{align}
    I_2(A:B)
    = S_A + S_B- S_{AB}
    =\frac{1}{2}
\ln{\frac{\text{det}\Theta_A \text{det}\Theta_B}{\text{det}\Theta}}
\end{align}
in terms of the R\'enyi-2 entropy. 
This mutual information 
is a measure of the total correlation between the two oscillators.  

Fig.~\ref{fig:3}a shows steady-state mutual 
information, $I_2(A:B)$, for various 
cases where there is a thermal bias 
between the two oscillators.  
$I_2(A:B)$ diverges 
as $\xi$ approaches the 
limit imposed by Eq.~\ref{eq:19}.
Since the fully synchronized (or anti-synchronized) steady-state is only 
achieved when $\xi = \pm 1$, a thermal 
bias will inhibit the formation of this state.
Fig.\ref{fig:3}(b) illustrates the
case where the two oscillators are at a common 
temperature.  Here, $I_2(A:B)$ diverges 
at $\xi=\pm 1$.  This means
that in the fully synchronized limit, 
a measurement on subsystem $A$ gives you 
full access to the information in subsystem $B$.

\subsubsection*{Discord}
For the given bi-partite system and the specific form of the covariance matrix, it is possible to filter out the classical part of the total mutual information as we showed in our previous work \cite{JPCL2024}. Following the work by Adesso, Datta et al.\cite{PhysRevLett.109.190502,PhysRevLett.105.030501} we can define the measurement-dependent classical part of the total mutual information by $\mathcal{J}_2(\rho_{A|B})\ge 0$. This leaves us with a lack of information about subsystem A when a Gaussian measurement is performed on subsystem B. Such measurements preserve the Gaussian nature of states. We have 
\begin{equation}
    {\cal J}_2 (\rho_{A|B}) = \sup_{\Sigma^{\Pi}_B} \frac{1}{2} \ln \left(\frac{\text{det} \Theta_A}{\text{det} \tilde{\Theta}_A^{\Pi}}\right)
    \label{eq:27}
\end{equation}
where
\begin{equation}
    \tilde{\Theta}_A^{\Pi} = \Theta_A - \Gamma_{AB}\left(\Theta_B + \Sigma^\Pi_B\right)^{-1} \Gamma^T_{AB}
    \end{equation}
is the Shur complement of the $\Theta_A$ block
and $\Sigma_B^{\Pi}$ is the seed of a Gaussian measurement on B. 
Naturally then the {\it Discord}\cite{OllivierZurek:2002} or the degree of quantumness can be defined as 
the difference between the total 
correlation and the purely classical 
contributions\cite{OllivierZurek:2002}.
\begin{equation}
\begin{split}
    {\cal D}_2 &=  I_2(A:B) - {\cal J}_2 (A|B)\\
    &=\inf_{\Sigma_B^{\Pi}}\frac{1}{2} \ln \left(\frac{\text{det}\Theta_B \text{ det}\tilde{\Theta}_A^{\Pi}}{\text{det}\Theta}\right).
    \label{eq:discord}
\end{split}
\end{equation}

In Ref.~\citenum{JPCL2024}
we showed that one can obtain a tight lower bound on the amount of quantum correlation 
between subsystems
by simply removing the 
coherences between the two subspaces within the 
total density matrix and computing the mutual 
information between the 
two remaining diagonal blocks. 
For Gaussian systems, this follows from Eq.~\ref{eq:27} since ${\cal J}_2(\rho_{A|B})$ is extremized under this condition and 
Eq.~\ref{eq:discord} becomes equal to the 
total correlation.

In our previous work on 
coupled qubit dimers, we found that depending upon the correlation 
within the environment as specified by $\xi$, by and large, coupling to a common reservoir 
introduces {\em quantum}
correlations between 
the spins.\cite{PMC2024,JPCL2024} 
For the case at hand, 
we have two possible sources of correlation
in our model. First, the 
direct coupling introduces correlation when the two oscillators are at different temperatures. Secondly, 
even when the direct coupling is turned off
or the two oscillators are at the same temperature, 
correlations within the 
common environmental mode
introduces correlations
between the two oscillators. Our results are
consistent with the formal results 
in Ref. 
\citenum{PhysRevLett.109.190502}
which show that  
 the only two-mode Gaussian states with zero Gaussian quantum discord are product states, 
 \textit{i.e.} states with no correlations at all, that constitute a zero measure set. 
 Consequently,  all correlated two-mode Gaussian states have nonclassical correlations certified by a nonzero quantum discord.

\section{Discussion}

We demonstrate here that synchronization 
and entanglement is a universal phenomina that 
can arise via {\em indirect} 
coupling  to a correlated
dissipative environmental 
mode. Environmental 
correlation imposes a 
symmetry on the system
by selectively 
increasing the 
relaxation rates for modes
that have the same
symmetry as the environmental
correlations leaving the remaining modes {\em effectively decoupled} from the environment. 
This selectivity will persist
in multi-mode systems, especially in system with 
a high degree of topological 
symmetery. One can use this 
to engineer dissipation and 
decoherence free domains within the total Hilbert 
space of the system. 
An example of enhanced collective behavior 
is the recent report of quantum 
collective motion in a superconducting circuit 
optomechanical platform composed of six-equivalent oscillators
in a hexagonal arrangement.\cite{Chegnizadeh:2024aa}
While the authors do not consider the influence of the
environmental noise, the experiments do suggest that 
similar devices could be fabricated that could have 
controllable degrees of environmental correlations. 

Our results also have implications in terms of
understanding the underlying
mechanisms for the transport
of quantum excitations in 
photosynthetic systems
in which observed long-lived exciton coherences are linked to vibronic correlation effects. The enhanced excitonic coherence has a profound influence on the enhancement of exciton transport and delocalization~\cite{engel2007evidence,panitchayangkoon2010long,chenu2015coherence,scholes2017using,Zhu:2024aa,Du:2021aa}.

One of the biggest challenges 
in quantum computing 
is decoherence—the loss of quantum coherence due to interactions with the environment. Quantum states, particularly superpositions and entanglement, are very fragile and prone to disruption due to 
environmental fluctuations.
This work suggests that by carefully
engineering the local environment
about each qubit and correlations between  qubits, one can construct
symmetry-adapted baths
that force the system into phase-locked state. For instance, synchronized quantum oscillators may continue to operate in coherence for longer durations, increasing the reliability of quantum computations and preserving the integrity of qubits.
By actively mitigating decoherence, facilitating error correction, and improving qubit control, synchronization could address several of the most pressing challenges in quantum computing. Additionally, it enables more efficient quantum information transfer, helps to stabilize entanglement by preserving fidelity\cite{PMC2024}, and supports the development of precise quantum clocks. These advancements will be pivotal in realizing robust and scalable quantum computers.

\section*{Acknowledgments}
This work was funded by the National Science Foundation (CHE-2404788) and the Robert A. Welch Foundation (E-1337).

\section*{Author contribution 
statement}
The authors acknowledge discussions with Prof. Carlos Silva (U. Montreal)
and Andrei Piryatinski (LANL), which led to the development of this model and series of calculations.  
Both authors discussed the results and contributed to the final manuscript.

\section*{Data Availability Statement}
Data files and source codes used in this work are freely and publicly available on the Wolfram cloud at the URL:
\url{https://www.wolframcloud.com/obj/a7bd6168-46c4-47e6-9f01-1755e79553f6}

\section*{Competing Interests Statement}
The authors have no competing interests to declare.


\end{document}